\documentclass[a4paper,11pt]{article}
\pdfoutput=1
\usepackage{amssymb,amsmath,amsfonts,makeidx,placeins,pbox,multirow}
\usepackage{graphicx,rotate,subcaption,color,slashed,cite,caption,epstopdf,verbatim}
\usepackage{array}
\usepackage{pstricks}
\usepackage{color}
\usepackage{amsmath}
\usepackage{mathtools}
\usepackage{amssymb}
\usepackage[normalem]{ulem}

\usepackage{MnSymbol}
\usepackage{tikz}
\usepackage{amssymb}
\usepackage{pifont}

\newcolumntype{P}[1]{>{\centering\arraybackslash}p{#1}}
\usepackage[colorlinks=true,
            linkcolor=magenta,
            urlcolor=blue,
            citecolor=blue]{hyperref}
\evensidemargin=\oddsidemargin
\def\to {\rightarrow}

\def\bea {\begin{eqnarray}}
\def\eea {\end{eqnarray}}

\def\barr{\begin{array}}
\def\earr{\end{array}}
\def\to{\end{rightarrow}}

\def\gev{\ensuremath{\mathrm{Ge\kern -0.1em V}}}

\begin{document}
\begin{center}
{\Large \bf Veltman Criteria in Beyond Standard Model Effective Field Theory  \\
\vspace{0.3cm}
of Complex Scalar Triplet} \\
\vspace*{0.5cm} {\sf  $^{a}$Jaydeb Das \footnote{jaydebphysics@gmail.com }, $^{b}$Nilanjana Kumar\footnote{nilanjana.kumar@gmail.com}} \\
\vspace{10pt} {\small }  {\rm  $^{a}$ Department of Physics and Astrophysics, University of Delhi, Delhi-110007, India \\ $^{b}$Centre For Cosmology and Science Popularization (CCSP), \\ SGT University, Gurugram, Haryana-122006, India} \\
\normalsize
\end{center}
%==========================================================================================================
\bigskip
%===========================================================================================================
\begin{abstract}
The Higgs mass is not protected by any symmetry in the Standard Model. Hence, the 
self-energy corrections to the Higgs mass become
large due to the quadratic divergence terms.
Veltman condition (V.C.) ensures that the coefficient of the 
quadratic divergent term either vanishes or becomes negligible. 
%If the Standard Model 
%is valid upto a scale ($\Lambda$) and new physics takes over beyond that; 
%Veltman condition demands $\Lambda\lesssim 760$ GeV. 
The non-observation of new physics has pushed the new physics scale
to be larger than 1 TeV, making it impossible to satisfy the Veltman condition in the Standard Model without 
very large fine-tuning.
Many attempts are made to satisfy the V.C. in Beyond Standard Model theories, 
%such as the
%Standard Model extended with a complex scalar triplet ($Y=1$). In these models, 
but the V.C. is hard to 
achieve at a very large scale ($\Lambda$).
Alternatively, it is possible that the new physics 
appears much above the Electroweak scale, and the effect of the new physics is observed 
in terms of the Wilson coefficients of the
Standard Model Effective Field Theory (SMEFT) operators. The V.C. can be addressed 
in the SMEFT framework.
%These SMEFT operators are obtained by 
%integrating out the degrees of freedom of the new physics above $\Lambda$.
In this paper, some specific new physics scenarios are considered at a very large scale. 
Below that scale, the effect of the new physics is observed as Beyond Standard Model Effective Field Theory (BSM-EFT).
We particularly study the type-II seesaw model with the complex scalar triplet ($Y=1$) in the context of V.C. 
We found that this particular model is the minimal model to generate all SMEFT operators 
that appear in V.C. and satisfies V.C.
We also examine the model parameter dependence of the Wilson coefficients in detail and show how the 
cancellation of the Wilson coefficients is highly dependent on some 
specific values of the model parameters.
\end{abstract}
%==========================================================================================================
%\tableofcontents
%=============================================
\section{Introduction}
%============================================

The smallness of the observed Higgs mass is confirmed by the experiments ~\cite{ATLAS:2012yve,CMS:2012qbp} 
at the Large Hadron Collider (LHC). However,
in the Standard Model (SM) of particle physics, the scalar mass (mass of Higgs boson) is not protected by any symmetry.
Hence, if SM is valid up to a large scale, Planck scale, the Higgs mass suffers 
from quadratic divergence ($\sim\Lambda^2$). To ensure that the mass of 
the Higgs boson is small, one has to consider a very large fine-tuning in the SM.
A way to ensure that the Higgs mass does not get large correction 
at a higher scale is coined as Veltman condition (V.C.)~\cite{veltman}. V.C. checks   
if the sum of all quadratically divergent 
terms coming from the self-energy diagrams of the Higgs boson is either zero or very small.
However, experiments such as the Large Hadron Collider (LHC) 
is pushing the New Physics (NP) scale towards $>1$ TeV, thus Veltman 
condition is not possible to satisfy in the SM, as it demands $\Lambda$ to be less than 760 GeV~\cite{Chakraborty:2014xqa}.

Simple extensions of SM have been studied in the literature 
\cite{Kundu:1994bs,grzadkowski,drozd,Chakraborty:2012rb,Chakraborty:2014oma,Chakraborty:2014xqa,Chakraborty:2016izr,Habibolahi:2022rcd,Decker:1979cw}, 
where the V.C. is valid but only in some regions of parameter space. 
Overall, there are two main concerns in these models: (1) These theories encounter different 
problems at a large scale, such as the potential becomes unstable leading to the invalidity 
of the theory beyond that scale, (2) The non-observation 
of the Beyond Standard Model (BSM) particles has pushed their masses above TeV scale~\cite{Workman:2022ynf}.

One may assume that SM is valid up to a certain scale ($\Lambda$) and above that
scale, some unknown symmetry appears to protect the Higgs mass, 
then the Higgs mass can be stabilized and the fine-tuning problem can be addressed. 
For example, in the Composite Higgs 
Scenario~\cite{Contino:2010rs}, where the Higgs is dissolved in higher degrees of freedom above the 
symmetry breaking scale or in Supersymmetric theories~\cite{Fayet:1976et}, where the 
bosonic and fermionic degrees of freedom cancel out exactly -- the Higgs mass 
is maintained to be finite and small. These theories also can not avoid 
a certain amount of fine-tuning \cite{Barnard:2017kbb,vanBeekveld:2019tqp} coming from several sources.
However, experiments are yet to confirm the existence of these theories.

These observations raise the question that what if the new physics lies at a very large scale.
In such a scenario, SM can emerge as an Effective Field Theory (SMEFT)~\cite{Brivio:2017vri} 
by integrating out the dynamics of the larger theory. The information of the heavy 
particles appearing in the loop is absorbed in the higher dimensional operators in the Effective Field Theory (EFT) and 
the theory is invariant under SM symmetries. 
Ref~\cite{Biswas:2020abl} has shown that the V.C. can 
be satisfied in the SMEFT framework by including 
the higher dimensional operators and their Wilson coefficients. Only a few of the operators are
relevant to the V.C. and they play a major role in satisfying the V.C.

In this paper, we take one step forward and ask this question what if 
the theory at a very high scale ($\Lambda$) includes specific type of BSM scenarios? and 
how that affects the V.C.? We adopt the 
Beyond Standard Model Effective Field theory (BSM-EFT) \cite{Adhikari:2020vqo} 
approach, which has been studied previously in Ref~\cite{Adhikari:2020vqo,Karmakar:2019vnq,Alanne:2017oqj,Bar-Shalom:2018ure,Du:2022vso,Li:2022ipc,Zhang:2021jdf,Crivellin:2016ihg}. 
In BSM-EFT the Lagrangian becomes invariant under the particular BSM model in consideration.
The motivation to study the V.C. in the BSM-EFT framework is twofold. 1) We can specifically check how 
many SMEFT operators are allowed by the model.
2) As the Wilson coefficients can be expressed in terms of the model parameters, the 
sign of the W.C., which is crucial to obtain V.C., comes naturally.
For calculation of the SMEFT operators we choose WARSAW basis~\cite{Grzadkowski:2010es,Jenkins:2013zja}. 
 
We begin with simple BSM scenarios (in BSM-SMEFT) such as scalar singlets, doublets and triplets (real or complex) 
and found that V.C. can be satisfied in all these models. We also found that 
the complex scalar triplet model with $Y=1$ is the minimal model (with only of one type of BSM particle) where 
it is possible to generate all the SMEFT operators that
contributes to the V.C. In Section 4 we discuss more on this \footnote{Type I and Type II seesaw models 
does not generate all SMEFT operators to satisfy V.C. Moreover a recent study~\cite{Du:2022vso} has also 
shown that these models are also not favored from the fact that 
the radiative electroweak symmetry breaking can not be triggered even at the Planck scale.}.  
Moreover, this particular model is well motivated in literature from other aspects as well:
1) Neutrino mass generation through the see-saw mechanism \cite{type2}, 2) type-II Leptogenesis scenario \cite{leptogenesis}
3) Enhancement of the $h\rightarrow \gamma\gamma$ branching ratio \cite{arhrib} etc, among many other\cite{logan,Ashanujjaman:2022ofg,Ashanujjaman:2021txz}.

In Section 2, we show how the V.C. depends on the SMEFT operators in the WARSAW basis~\cite{Grzadkowski:2010es}.
In Section 3, we discuss some specific models in the BSM-EFT scenarios, and express the 
Wilson coefficients of $Y=1$ complex scalar triplet model in terms of the model parameters. In Section 4, 
we show how V.C. is achieved by the exact cancellation 
of the Wilson coefficients at different scale. We also interpret the result in terms of the model parameter space.
Then in Section 5, we conclude.
%while satisfying the constrains from EW observables(), Neutrino data(), Collider data(), 
%and also stability and unitarity conditions.
%One the one hand, from the point of view of naturalness, 
%it is known that the type-I and -III seesaw models would live above
%the GUT scale [25–27], making them impossible to be produced or tested directly at current
%experiments. 
%Whereas,in type-II seesaw model the neutrino Yukawa couplings could be of O(1) 
%with a relatively light triplet at the O(1 TeV) scale as
%long as the triplet vacuum expectation value (vev) is agnostically small. The masses of 
%the BSM particles are also testable at
%colliders Refs. [28–38].

%We have considered the running of SM parameters and also the running of the Wilson coefficients.
%Here we have considered the constrains from the electroweak precession data, 
%neutrino mass while evaluating the model parameters at $\Lambda$.
%the small neutrino mass and $\rho=1$both demands the vev of the triplet to be very small 
%or MAss of the triplet to be very large.
%We present three BP scenario, 1) 10 TeV \footnote{Not importantt from VC point of view but testable in near future},
% 100 TeV and 3) $10^6$ TeV.
%This model also predicts degenerate and non degenerate spectrum, based on the values of the triplet VEV.
%THe current limit on the mass difference is $\Delta M < 40 GeV$.
%2201.05082
%2201.04646
%===============================================
\section{SMEFT operators and Veltman Condition}
%===============================================
The physical mass of the Higgs in the Standard Model can be written in terms of the
bare mass term $m_h(0)$ and the higher-order self-energy corrections:
\begin{equation}
m^2_h = m^2_h(0) + \delta m^2_h = m^2_h(0) + \text {Log Div. Term + Quadratic Div. Term + Finite terms},
\end{equation}
\\
where the assumption is that the SM is valid up to the scale $\Lambda$ and the 
correction terms are coming from the loop diagrams 
involving scalars, fermions and bosons in the loop. 
The $d=4$ potential in the Standard Model in terms of Higgs doublet ($H$) is 
\begin{equation}
V(H) = -m_{H}^2 H^\dag H + \lambda (H^\dag H)^2 .
\end{equation}
This leads to the correction to the higgs mass and the quadratic divergent contribution can be expressed as, 
\begin{eqnarray}
(\delta m_h^2 )_{\rm SM}& =& \frac{\Lambda^2}{16 \pi^2}\left( 6 \lambda + \frac{9}{4}g_W^2 + \frac{3}{4} g_Y^2 - 6 y_t^2 \right),
\end{eqnarray}
where, $g_Y$ and $g_W$ are the $U(1)_Y$ and $SU(2)_L$ gauge couplings respectively and 
$g_t = 2m_t/v$ is the top quark Yukawa coupling. Here we neglect the couplings of 
the lighter quarks and $\Lambda$ is the cut-off scale.
The Veltman condition (V.C.) demands that $\delta m_h^2\sim 0$ or at least controllably small. 
With the observed Higgs mass at 125 GeV, the condition to 
make $\delta m_h \sim 0$ demands $\Lambda < 760$ GeV, which is already ruled out by LHC.
One way to solve this problem is to introduce new particles, which 
can contribute in the loops and soften the fine-tuning by ensuring 
exact cancellation or partial as we have already discussed in the introduction.

A popular way to address this problem is to consider the effects of the 
higher dimensional operators in the EFT framework. Let us assume that the New Physics (NP)
exists at a very high scale $\Lambda$. The effect of NP can be integrated out 
at $\Lambda$ and this will effectively give us SM,
plus some effective operators involving only the SM fields. This is known as the 
Standard Model Effective Field Theory (SMEFT)~\cite{Brivio:2017vri}.
The Lagrangian, which incorporates dimension six SMEFT operators in addition 
to the Standard Model dimension four operators, can be expressed as, 
\begin{equation}
\mathcal{L} = \sum_i \mathcal{C}_{4i}\mathcal{Q}_{4i} +\frac{1}{\Lambda^2}\sum_{i} \mathcal{C}_{6i}\mathcal{Q}_{6i}.
\end{equation}
In contrast to $\mathcal{C}_{4i}$, which is the only function of the parameters linked to the degrees of freedom in the Standard Model, $\mathcal{C}_{6i}$ are the Wilson coefficients, which are functions of the integrated out dynamics at $\Lambda$. 
These operators can be expanded at any choice of basis, for example,
HISZ basis \cite{Hagiwara:1996kf,Brivio:2021alv}, Warsaw basis~\cite{Grzadkowski:2010es,Jenkins:2013zja}, 
SILH basis~\cite{Giudice:2007fh} etc. 
The set of dimension six operators that involves Higgs in Warsaw basis are:
\begin{eqnarray}\label{eq:smefts}
&\mathcal{Q}_H = (H^\dagger H)^3,\,\,\ \mathcal{Q}_{HD} = (H^\dagger\mathcal{D}_\mu H)^\ast(H^\dagger\mathcal{D}^\mu H),\,\,\ \mathcal{Q}_{H\square} = (H^\dagger H)\square (H^\dagger H) \nonumber\ \\
 &\mathcal{Q}_{HB} = (H^\dagger H) B_{\mu\nu}B^{\mu\nu},\,\,\ \mathcal{Q}_{HW} = (H^\dagger H)W_{\mu \nu}^aW^{a,\mu \nu} ,\,\,\
 \mathcal{Q}_{GG} = (H^\dag H) G^A_{\mu \nu}G^{A,\mu \nu} \nonumber\ \\
&\mathcal{Q}_{HWB} = (H^\dag \tau_a H)B^{\mu\nu}W^a_{\mu\nu}
\end{eqnarray}
It can be shown that the last operator, $\mathcal{Q}_{HWB}$ does not contribute 
Higgs self-energy correction\cite{Biswas:2020abl}. The first operator, $\mathcal{Q}_{H}$ will also not 
contribute at one-loop level as the Higgs does not develop a vev at $\Lambda$.
There can be the appearance of the operators 
involving the gluons of the form $\mathcal{Q}_{GG} =( H^\dag H) G^A_{\mu \nu}G^{A,\mu \nu}$.
However, while considering BSM-EFT framework with heavy scalars, this operator does not 
contribute as scalars do not carry any color charge. Note that, these operators can be 
written in any basis, for example Ref~\cite{Biswas:2020abl} choose the HISZ basis. 
We choose the Warsaw basis because it is self consistent at one loop~\cite{Jenkins:2013zja,Alonso:2013hga} and 
easier to check the running of the Wilson coefficients in Warsaw basis.

The correction to the Higgs mass from the higher order terms in the Lagrangian
is given by 
\begin{eqnarray}\label{eq:totmass}
(\delta m_h^2)_{\rm total} & =& \frac{\Lambda^2}{16 \pi^2} \sum_i f_i( \mathcal{C}_{4i},\mathcal{C}_{6i}) +\frac{\Lambda^2}{(16 \pi^2)^2} \sum_i g_i (\mathcal{C}_{4i},\mathcal{C}_{6i}) 
\end{eqnarray}
Here $f_i$ and $g_i$ are one loop and two loop correction to the Higgs mass.  
The V.C., $\delta m_h^2\sim 0$ translates into \\
\begin{equation}  
f(\mathcal{C}_{4i},\mathcal{C}_{6i}),g(\mathcal{C}_{4i},\mathcal{C}_{6i})\sim 0
\end{equation} 
if loop contributions are considered separately.
The coefficients, $\mathcal{C}_{4i}$ and $\mathcal{C}_{6i}$ are function of $\Lambda$ and the BSM model parameters.
Hence Eq:\ref{eq:totmass} can be written in terms of the SM and higher dimension operators contribution as, 
\begin{eqnarray}
(\delta m_h^2)_{\rm total}  \equiv (\delta m_h^2 )_{\rm SM} \big(f_i(\mathcal{C}_{4i}),g_i(\mathcal{C}_{4i})\big)+(\delta m_h^2 )_{\rm HO} \big(f_i(\mathcal{C}_{6i}),g_i(\mathcal{C}_{6i})\big)
\end{eqnarray}

Also, it has been shown in Ref~\cite{Biswas:2020abl}
that at $d\geq 8$, the SMEFT operators are not able to produce any $\Lambda
^6$ divergence, which 
will produce any effective $\Lambda^2$ divergence while calculating the self-energy correction of Higgs mass.
There are studies in the literature, where the V.C in terms of EFT has been studied in detail
\cite{Biswas:2020abl,Passarino:2021uxa,Abu-Ajamieh:2021vnh}.
In particular it has been shown in Ref:~\cite{Biswas:2020abl} that it is possible 
to satisfy the V.C for appropriate values 
and sign of the Wilson coefficients at large $\Lambda$.
%==============================================================
\section{BSM-EFT with Complex Scalar Triplet}
%==============================================================
%The potential in the SM is given by,
%\begin{eqnarray}
%V &=-\mu_H^2 (H^+ H) + \Lambda_H (H^+ H)^2& 
%\end{eqnarray}
%LHC puts a limit on new physics beyond 1 TeV in most of the cases, 
%showing no direct evidence of new physics beyond SM.  
%Hence, it might be possible that the new physics manifests itself 
%at a very large scale ($> 100$ TeV ) or so. Hence one can assume a particular 
%theory beyond that scale and study its effect via the effective theory framework.
%The BSM particles are very heavy and their effects are integrated out at $\Lambda$.
%This way, by assuming a particular theory which is motivated otherwise, 
%we get a smaller subspace spanned by SMEFT coefficients compared to full SMEFT. 
%Another advantage is that the Wilson Coefficients can be determined in 
%terms of the model parameters. hence one can constrain the 
%model parameters by placing limits to the Wilson Coefficients.
In the above section, we saw that only four operators in the WARSAW basis are involved in the V.C.
Now, we assume that the new physics at a large scale follow certain symmetries of a BSM model which effectively 
produces SM as an EFT. It has been already shown in literature how some BSM extensions
\cite{Kundu:1994bs,grzadkowski,drozd,Chakraborty:2012rb,Chakraborty:2014oma,Chakraborty:2014xqa,Chakraborty:2016izr,Habibolahi:2022rcd,Decker:1979cw}
address the V.C. In this paper we consider them to appear at a large scale and dictate the underlying symmetry of the EFT. 

In this BSM-EFT framework, these 4 operators may or may not be 
possible to generate at one loop, depending on the underlying symmetry of the model at scale $\Lambda$.
In Table:~\ref{tab:1}, we present if these 4 operators can be generated at one loop in 
some simple BSM-EFT cases with additional scalar(s) or not 
\footnote{Note that we are not checking non scalar extensions of SM because, the sign of the top-loop 
contribution (dominant contribution) or rather fermionic contribution is opposite to the other diagrams 
with a gauge boson or a scalar in the loop. Therefore, V.C. is hard to solve 
by adding non scalar particles such as vector-like quarks or fermions, additional gauge bosons etc.}.
For the calculation, we have implemented the Lagrangian of each model in CoDEx~\cite{DasBakshi:2018vni,Anisha:2021hgc}
and generated the Wilson coefficients.~\footnote{We have also cross checked our result with Matchmakereft\cite{Carmona:2021xtq}.}.
%----------------------------------------------------
\begin{table}[!h]
  \begin{center}
\begin{tabular}{|c|c|c|c|c|c|}
\hline
\hline
Model& Quantum No & $\mathcal{Q}_{HD}$ & $\mathcal{Q}_{HB}$ & $\mathcal{Q}_{HW}$ & $\mathcal{Q}_{H\square}$ \\
\hline
Real Scalar Singlet &(1,1,0) & \ding{51}  & \ding{55} & \ding{55} & \ding{55}\\
Real Scalar Triplet &(1,3,0)&\ding{55}&\ding{55}&\ding{51} &\ding{51} \\
Complex Scalar Triplet &(1,3,1)&\ding{51} &\ding{51} &\ding{51} &\ding{51} \\
\hline									
Complex scalar doublets (2HDM) &(1,2,$\pm$1/2)&\ding{51} &\ding{55}&\ding{51} &\ding{51} \\
\hline
Real Scalar Singlet + &(1,1,0)& \ding{51} &\ding{55}&\ding{51} &\ding{51} \\
Real Scalar Triplet &(1,3,0)&&&&\\
\hline
Complex Scalar Triplet + &(1,3,1)&\ding{51} &\ding{51} &\ding{51} &\ding{51} \\
Complex Scalar Doublet& (1,2,1/2) &&&&\\
\hline
\end{tabular}
\end{center}
    \caption{\label{tab:1}\em SMEFT operators in WARSAW basis in different BSM-EFT scenarios.}
\end{table}
%---------------------------------------------------------------

Among all popular models, we have found that 
BSM-EFT with complex scalar triplet is the minimal model 
where all four Wilson coefficients are generated at one loop.
In other models, the number of operators is less than four except 
for the model with complex scalar triplet with the additional doublet.
In 2HDM scenario and real scalar singlet + triplet model, only three operators can be generated, whereas, 
in the complex scalar singlet model, only 2 operators are generated.
The real scalar singlet model generates only one operator. All this models, with four or less number of 
operators satisfies the V.C. But, in is intuitive to see that the parameter space of the W.C.s is 
more constrained in a model which generates less than four of EFT Operators. Hence, we 
chose to study the complex scalar triplet model in detail as 
the {\it minimal model} (with only one type of BSM particle) along with other 
motivations as mentioned in the Introduction.

Let us consider that beyond the scale $\Lambda$, there exists a heavy complex 
triplet, $\Delta$, with weak hypercharge $Y=1$.
The most general renormalizable tree-level scalar potential of such a model is given by
\begin{eqnarray}
V(H	, \Delta) &=& - m_H^2 \left(H^\dag H\right) + M^2 \mathrm{Tr} \left[\Delta^\dag \Delta\right] + \left( \mu_\Delta H^T i \sigma^2 \Delta^+ H +h.c  \right) + \lambda \left(H^\dag H \right)^2 + \lambda_1 \left(H ^\dag H \right) \mathrm{Tr} \left[\Delta^\dag \Delta\right] \nonumber\\\
&+&  \lambda_2  \left(\mathrm{Tr} [\Delta^\dag \Delta]\right)^2 + \lambda_3 \mathrm{Tr} \left[(\Delta^\dag \Delta)^2\right] + \lambda_4 \left(H^\dag \Delta \Delta^\dag H\right).
\end{eqnarray}
The extra Yukawa term for neutrino mass generation is, 
\begin{eqnarray}
\mathcal{L}_Y &=& y_\Delta \ell ^T i C i \sigma^2 \Delta \ell + h.c.
\end{eqnarray}
Here the trilinear coupling $\mu_\Delta$ can be taken as positive by absorbing its phase into $H$ and $\Delta$.
The total Lagrangian is,
\begin{equation}
\mathcal{L} =\mathcal{L}_Y - V(H	, \Delta),
\end{equation}
The details of this model is summarized in `Model Description' section of the Appendix.

We explicitly show the expansion of the dimension six operators (as listed in~Eq:\ref{eq:smefts}), in the `Calculation' section of the Appendix.
The Higgs mass correction in terms of W.C.s is WARSAW basis\footnote{ Please check to the Appendix for the result in different basis.} is given by,  
\begin{eqnarray}
(\delta m_h^2)_{\rm BSM} &=& \frac{\Lambda^2}{16 \pi^2}\Big( -3 \mathcal{C}_{HD} + 12 \mathcal{C}_{H\square} + 9 \mathcal{C}_{HW} + 3 \mathcal{C}_{HB} \Big)  \nonumber\ \\
&+&
\frac{\Lambda^2}{(16 \pi^2)^2}\left(54 \mathcal{C}_H -\frac{9}{2}(g_Y^2 + 3 g_W^2)\mathcal{C}_{HD} + 108 g_W^2 \mathcal{C}_{HW} \right).
\end{eqnarray}
The total correction to the Higgs mass is, 
\begin{eqnarray}
\delta m_h^2= (\delta m_h^2)_{\rm SM} +(\delta m_h^2)_{\rm BSM}.
\end{eqnarray}

The Wilson coefficients appearing in one loop contribution 
in $(\delta m_h^2)_{\rm BSM}$ can be expressed in terms of the model parameters as:
\begin{eqnarray}
\mathcal{C}_{HD} &=& -\frac{g_Y^4}{320 \pi ^2}+\frac{4 \mu_\Delta^2}{M^2}-\frac{\lambda_4^2}{24 \pi ^2}+\frac{11 g_Y^2 \mu_\Delta^2}{24 \pi ^2 M^2}-\frac{8 \mu_\Delta^4}{3 \pi ^2 M^4}+\frac{\lambda_4 \mu_\Delta^2}{6 \pi ^2 M^2}+\frac{3 \lambda \mu_\Delta^2}{8 \pi ^2 M^2}\\
\mathcal{C}_{H\square} &= & -\frac{g_W^4}{1920 \pi ^2}+\frac{2 \mu_\Delta^2}{M^2}-\frac{\lambda_1^2}{16 \pi ^2}-\frac{\lambda_1 \lambda_4}{16 \pi ^2}-\frac{\lambda_4^2}{192 \pi ^2}-\frac{g_W^2 \mu_\Delta^2}{96 \pi ^2 M^2}+\nonumber\ \\
&+&
\frac{11 g_Y^2 \mu_\Delta^2}{96 \pi ^2 M^2}-\frac{49 \mu_\Delta^4}{12 \pi ^2 M^4}+\frac{\lambda_1 \mu_\Delta^2}{8 \pi ^2 M^2}+\frac{\lambda_4 \mu_\Delta^2}{48 \pi ^2 M^2}+\frac{3 \lambda \mu_\Delta^2}{4 \pi ^2 M^2}\\
\mathcal{C}_{HB} &=& \frac{g_Y^2 \lambda_1}{32 \pi ^2}+\frac{g_Y^2 \lambda_4}{64 \pi ^2}+\frac{11 g_Y^2 \mu_\Delta^2}{64 \pi ^2 M^2}\\
\mathcal{C}_{HW} &=& \frac{g_W^2 \lambda_1}{48 \pi ^2}+\frac{g_W^2 \lambda_4}{96 \pi ^2}+\frac{25 g_W^2 \mu_\Delta^2}{192 \pi ^2 M^2}.
\end{eqnarray}
Here, $M$ is the mass of the heavy triplet. For the theory to be valid, it is sufficient to 
assume that $M$ is greater than $\Lambda$. We assume the order of magnitude to be the same 
for $M$ and $\Lambda$ in our calculation as a limiting scenario. For $M>> \Lambda$, the W.C.s will 
obtain smaller values.
%==============================================================
\section{Result}
%==============================================================
In order to satisfy the Veltman condition, we consider the one loop correction to the Higgs mass ($\delta m_h^2$) and 
fix two benchmark scenarios at $\Lambda=100$ TeV and
$\Lambda=10^6$ TeV. Then we figure out the model parameter space of $\lambda_1$
and $\lambda_4$, for which the quadratic divergence in $\delta m_h^2$ cancels out exactly.
The SM input parameters: $g_W$, $y_t$, $g_Y$ and $\lambda$ are determined at $\Lambda$
by solving the two loop Renormalized Group Equation (RGE)'s. 
$\lambda_1$ and $\lambda_4$ are varied in such a way that the 
Wilson coefficients obey the perturbative limit and the running of 
the Wilson coefficients from $\Lambda$ to the Electroweak scale are smooth. 
The values of the tree level couplings ($\lambda$ and $\mu_H$) also shift due to the higher dimensional operators.
The parameter $\lambda$ can not be more than $O(1)$ and this puts an upper limit 
on the quantity $\frac{\mu_{\Delta}^2}{2M^2}<O(1)$, where $\mu_\Delta=\frac{\sqrt{2}v_\Delta M^2}{v_H^2}$, in the limit of large scalar 
triplet mass. Also, precision measurements have set the value of the $\rho$ parameter 
to be in the range $1.00038 \pm 0.00020$ (ref). This constrains the {\it vev} of the triplet ($\Delta$) to be less than $2.56$ GeV\cite{Ghosh:2022zqs}.
%=======================
\begin{figure}[ht]
\begin{center}
\includegraphics[scale=0.38]{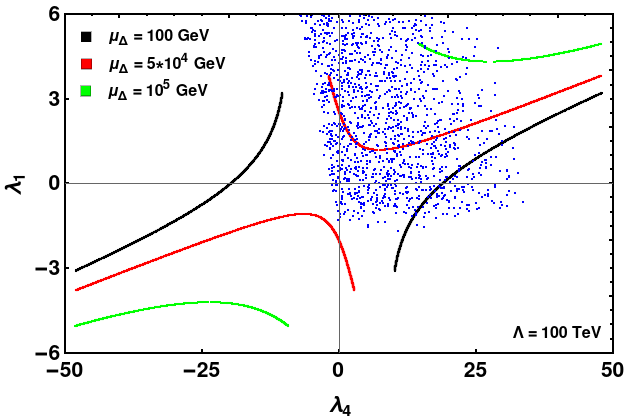}
\includegraphics[scale=0.38]{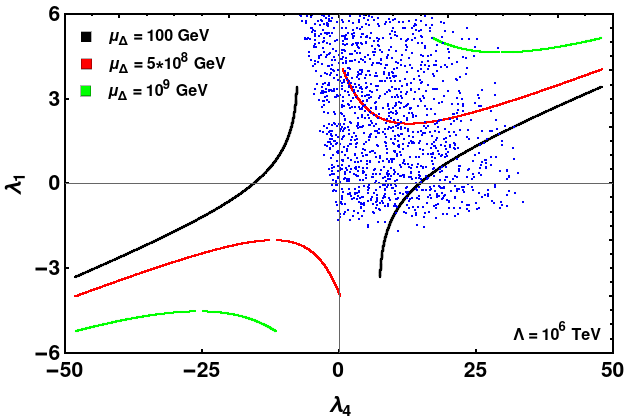}
\caption{Variation of $\lambda_1$ and $\lambda_4$ with $\mu_{\Delta}$ at two benchmark values of $\Lambda$. V.C. is satisfied over the lines. 
The blue points show the allowed region when the full theory with Triplet is bounded from below and satisfies perturbative unitarity conditions.}
\label{fig:fig1}
\end{center}
\end{figure}
%========================

In Fig:~\ref{fig:fig1} we show    
the parameter space of $\lambda_1$ and $\lambda_4$ which satisfies the V.C.
We found that both positive and negative values of $\lambda_1$ and $\lambda_4$ satisfy V.C.
The green line represents the highest possible value of $\mu_\Delta$, which comes from the constraint 
$\frac{\mu_{\Delta}^2}{2M^2}\sim O(1)$.
The nature of these plots is highly dependent on the values of $\mu_\Delta$, because 
the Wilson coefficients have $(\mu_\Delta/M)^2$ and $(\mu_\Delta/M)^4$
dependence with additional suppression of $1/16\pi^2$. 
The sign of the W.C's come naturally from the fact that they 
are determined in terms of the model parameters $\lambda_1$, $\lambda_4$ and $\mu_\Delta$, 
which we allow to vary freely with the above mentioned constraints.
We found that the V.C. is satisfied  
even if $\Lambda$ is very large ($10^6$ TeV).
It is also essential to check whether the full theory with the Triplet scalar is well behaved above $\Lambda$. For that
we have considered the vacuum stability, bounded from below and unitarity conditions \cite{Das:2016bir,Arhrib:2011uy} and the 
blue points in Fig:~\ref{fig:fig1} satisfy these conditions. 
We have found that the positive values of $\lambda_1$ and $\lambda_4$ 
are largely preferred for the full theory with the Triplet to be well behaved.
Hence V.C can be satisfied in a constrained region where the full theory with the triplet obeys stability 
and unitarity condition. 
The $\lambda_1$ and $\lambda_4$ parameter space remains mostly unchanged at large $\Lambda$ because 
the Wilson coefficients do not change much with $\Lambda$ (shown later). Although, a slight variation 
in the parameter space is present due to the running of SM parameters.
The cancellation in the V.C is dependent on the precision of the input parameters, which is also the source of negligible amount of fine tuning.
%\footnote{Note that, for other models, where the number of Wilson Coefficient is less than 4 can be generated, 
%the exact cancellation will be harder to obtain and the amount of fine-tuning will also be very high 
%compared to this model.}.
%=======================
\begin{figure}[h]
\begin{center}
\includegraphics[scale=0.38]{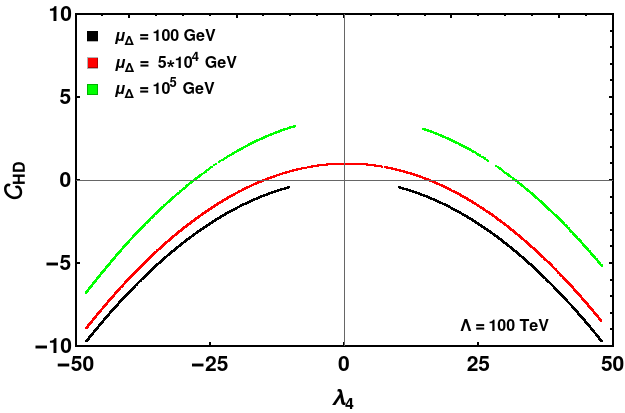}
\includegraphics[scale=0.38]{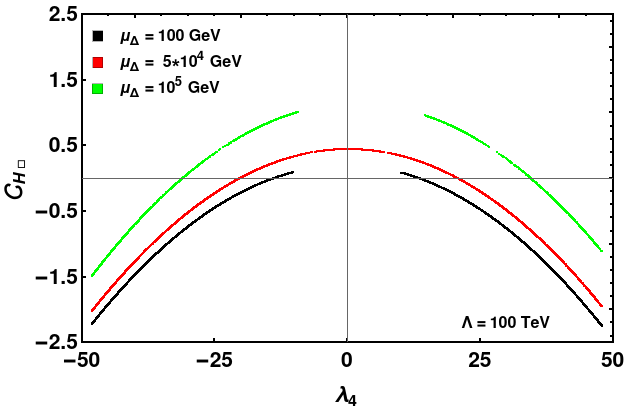}
\includegraphics[scale=0.38]{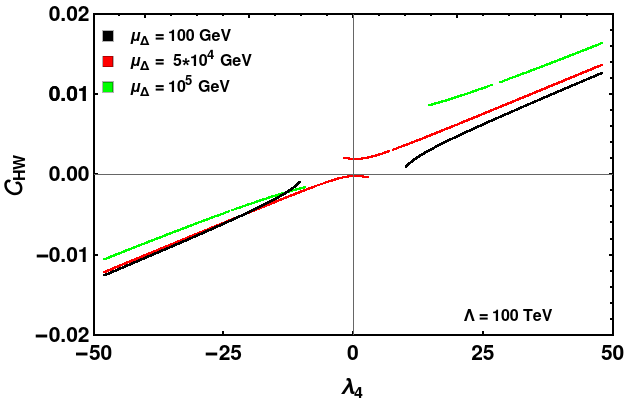}
\includegraphics[scale=0.385]{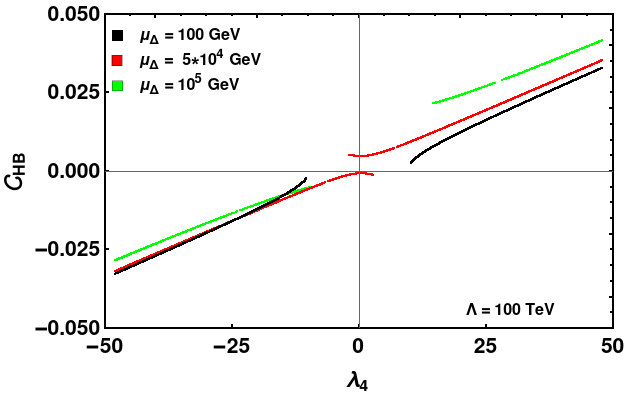}
\caption{Variation of the Wilson coefficients with model parameter $\lambda_4$ when V.C. is satisfied.}
\label{fig:fig2}
\end{center}
\end{figure}
%==================================

In Fig:~\ref{fig:fig2}, we show the variation of the Wilson coefficients with the 
model parameter $\lambda_4$ at 100 TeV. The corresponding value(s) of 
$\lambda_1$ can be inferred form Fig:~\ref{fig:fig1}.
The Wilson coefficients show similar behavior at the other benchmark scenario. 
For Wilson coefficients $\mathcal{C}_{HD}$ and $\mathcal{C}_{H\Box}$, negative values are more 
preferred, whereas, for $\mathcal{C}_{HW}$ and $\mathcal{C}_{HB}$, both positive and negative values are allowed
\footnote{The sign of the W.C. is different in different SMEFT basis. We list the transformation rules of the 
WC's in the Appendix.}.
However, when $\lambda_4$ is negative, almost all coefficients are negative,
except for some values of the $\mathcal{C}_{HD}$ and $\mathcal{C}_{H\Box}$. Again, when $\lambda_4$ is positive, 
$\mathcal{C}_{HW}$ and $\mathcal{C}_{HB}$ are always positive but $\mathcal{C}_{HD}$ and $\mathcal{C}_{H\Box}$ are mostly negative except 
for some values as shown in Fig:~\ref{fig:fig2}. Thus, it is visible that 
the cancellation among the Wilson coefficients is not {\it ad-hoc}, but is controlled by the model parameters.
We have checked the V.C n=by considering the two loop contribution to Higgs mass correction as well but 
due to the extra suppression by ($1/16\pi^2$), the effect is not visible.
%==================================
\begin{figure}[h!]
\begin{center}
\includegraphics[scale=0.405]{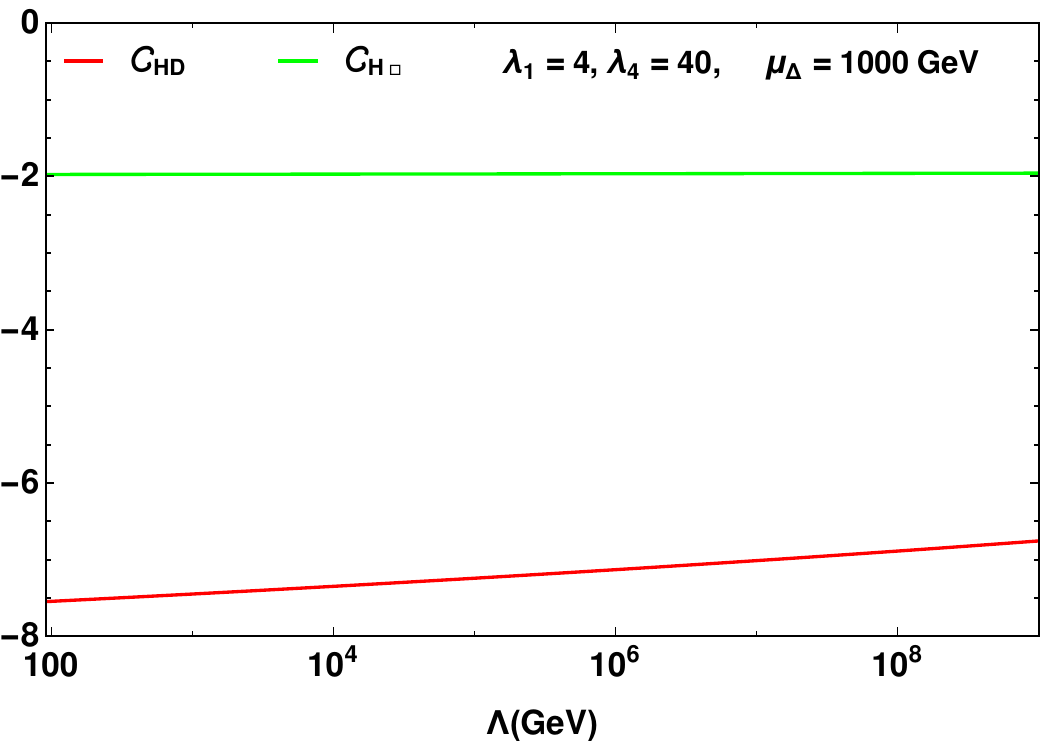}
\includegraphics[scale=0.42]{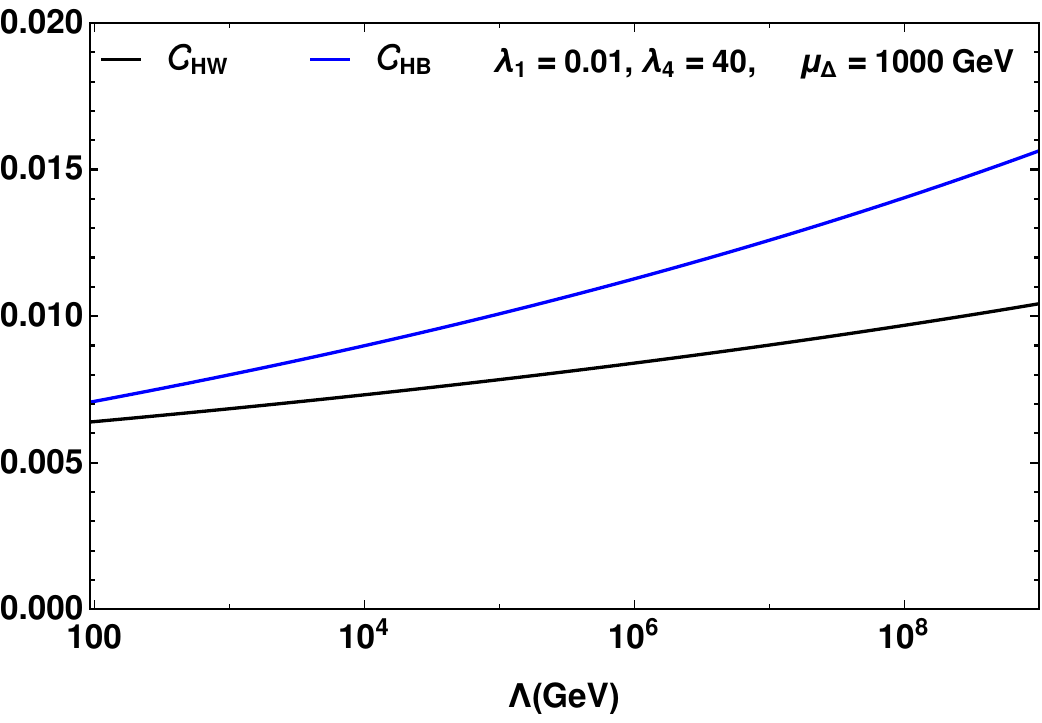}
\caption{Running of the Wilson coefficients from $\Lambda=10^6$ TeV to the cut-off scale for one set of model parameters.
We have kept the value of $\mu_\Delta$ to be fixed at $1$ TeV.}
\label{fig:fig3}
\end{center}
\end{figure}
%========================

We have also checked the running of the Wilson coefficients from the effective scale $\Lambda$ to the Electroweak scale.
We show the running of the Wilson coefficients in Fig:~\ref{fig:fig3} for a particular choice of the 
model parameters, $\lambda_1=4.0$ and $\lambda_4=40$. 
This particular choice of parameter represents the maximum
possible value of these parameters. 
We found that the values of these W.C.s do not change much and also the sign of the W.C.s do not change in the 
running.
The conclusion remains the same for other allowed values of $\lambda_1$ and $\lambda_4$ that satisfy the V.C.
The values of W.C.s ($C_i (1\rm TeV)^2/\Lambda^2$) are highly constrained at the EW scale~\cite{Ellis:2020unq} from 
various low energy experiments. 
The values of the W.C.s that satisfy the V.C., as shown in Fig:~\ref{fig:fig2}, Fig:~\ref{fig:fig3},  
lie within the current experimental bounds.
%================================================
%\paragraph{Constraints}
%================================
%\begin{itemize}
%\item The vev of the Higgs and the Triplet, $v_\phi$ and $v_\Delta$ satisfy 
%the condition,  $v^2=v_H^2+v_\Delta^2= (246)^2$ GeV 
%and $|\sin\alpha|\sim \frac{2v_\Delta}{v_H}$.
%The Higgs-to-diphoton decay rate at the LHC constrains $\alpha$ such that
%$|\sin\alpha|<0.3$ at 95~\% CL [1903.02493]. On the other hand, the most 
%recent precission measurements of the $\rho$parameter gives 
%$\rho = 1.00038 \pm 0.00020$, resulting in $v_\Delta <2.56<<v_{H}$.

%\item The neutrinos also get masses via 
%see-saw mechanism, 
%\begin{eqnarray}
%(m_\nu)_{\alpha\beta} = \sqrt{2} (y_\nu)_{\alpha\beta} v_\Delta, 
%\end{eqnarray}
%generated by the $d=5$ Weinberg operator. 
%The current limit on the sum of neutrino mass is $< 0.12$ eV as suggested by the 
%recent cosmological data from the Planck satellite. 

%\item Now, the tree level couplings ($\lambda$ and $\mu_H$) 
%also gets a shifted due to the oneloop matching at $\Lambda$.
%The parameter $\lambda$ can not be more than O(1) and this puts an upper limit 
%on the quantity, $\frac{\mu^2}{2M^2}<O(1)$, where 
%\begin{eqnarray}
%\mu_\Delta=\frac{\sqrt{2}v_\Delta M^2}{v_H^2}.
%\end{eqnarray}
%in the limit of large masses of the triplet ($M$).

%\item The mass difference among the charged and neutral components in this model is 
% $\delta m = \frac {\lambda_4 v_\Delta^2}{4M}$.
%The S, T and U parameters are sensitive to $\delta m$, and places a limit 
%$|\delta m| < 40$ GeV.
%\end{itemize}
%=================================================
\section{Conclusion}
The Veltman condition can not 
be satisfied within the framework of the Standard Model because of 
significant quadratic divergences to the Higgs self-energy correction 
if the cutoff scale $\Lambda$ is $\sim 1$ TeV or higher. 
In addition to the dimension four operators from the Standard Model, 
we have also included dimension six operators whose contributions to the 
Higgs mass correction results from integrating out the heavy triplet scalar 
with hypercharge. 
We show how the quadratic divergence of the Higgs 
self-energy vanishes in this particular model due to the cancellation 
among the SM parameters and the Wilson coefficients. 

We have shown the relevant SMEFT operators which contribute to the V.C and found agreement with Ref.~\cite{Biswas:2020abl}. 
The W.C. of the operators are
expressed in terms of the model parameters of the complex scalar triplet model. 
Hence, in this study the sign of the Wilson coefficients 
are not ad-hoc, it is driven by the larger theory, which is a heavy triplet scalar in our case.
We found that this model generates all four operators that appear in the V.C., allowing more 
parameter space to the W.C.s compared to the other models, where the number of operators is less than four.
However, the values of the Wilson coefficients will 
be different in every model, as it is controlled by the specific model parameters.

To achieve the Veltman condition, it should be noted that the contributions 
from two particular dimension six operators $\mathcal{Q}_{HD}$ and $\mathcal{Q}_{ H\square}$ play 
a dominating role in canceling out the quadratic divergences. 
We show the parameter space where V.C is satisfied for both positive and negative values of the model parameters.
We found that the V.C can be satisfied in a constrained region where the full theory with the triplet obeys stability 
and unitarity condition as well.  
We have observed that for energy scales $\Lambda$ = 100 TeV and $10^6$ TeV, the cancellation is almost 
similar. This is because the value of the W.C.s does not change much with $\Lambda$ and 
the insignificant amount of change appears due to the running of the SM parameters.
If we introduce some relaxation in the V.C., by allowing some amount of fine-tuning, the model parameter space 
will surely enlarge, but it will get narrower with the increasing values of $\Lambda$.
Thus, the Veltman condition can be easily satisfied in the framework of 
effective field theory, when a complex scalar triplet exists at a very large scale. 

The study of this model as an Effective Field Theory can also be useful to revisit the 
Type II leptogenesis scenario, where it will be possible to generate specific dimension six terms which 
are dictated by the model.

{\it \bf Acknowledgements}: JD acknowledges the Council of Scientific and Industrial Research (CSIR), Government of India, for the SRF fellowship grant with File No. 09/045(1511)/ 2017-EMR-I. JD also would like to acknowledge Research Grant No. SERB/CRG/004889/SGBKC/2022/04 of the SERB, India. The work of NK is supported by Department of Science and Technology, Government of India under the SRG grant, Grant Agreement Number SRG/2022/000363 and CRG grant with Grant Agreement Number CRG/2022/004120. We thank Prof. Anirban Kundu and Dr. Supratim Das Bakshi for the useful discussion during the preparation of the manuscript. We also thank the Referees for their valuable suggestions.
%===========================================================================================================
\section {Appendix}
%-------------------------------
\paragraph{\underline {Model Description}:}
%------------------------------
In the type-II seesaw model, the scalar sector is extended by a complex scalar triplet($\Delta$) with hypercharge $1$, 
in addition to the Higgs doublet ($H$). Explicitly the fields can be written as,
\begin{equation}
H\left(1,2,+1/2 \right) =
\begin{pmatrix}
\phi^+\\ \phi^0
\end{pmatrix}
,\,\,\
\Delta\left( 1,3,+1 \right) =
\begin{pmatrix}
\Delta^+/\sqrt{2}& \Delta^{++} \\ \Delta^0 & -\Delta^+/\sqrt{2}
\end{pmatrix}
\end{equation}
The numbers in the parentheses represent the charges of $SU(3)_C \times  SU(2)_L \times U(1)_Y $ gauge group of the SM. 
The neutral components are:
\begin{equation}
\phi^0= \frac{v_{H} +h +i\phi_3}{\sqrt{2}} ,\,\,\ \Delta^0 = \frac{v_{\Delta}+\delta+i\xi}{\sqrt{2}} 
\end{equation}
The kinetic terms corresponding to the scalar fields are given as
\begin{equation}
\mathcal{L}_{\rm kin}\supset (D^\mu H)^\dag D_\mu H + \mathrm{Tr} \left[ (D^\mu \Delta)^\dag (D_\mu \Delta)\right]	,
\end{equation} 
with the covariant derivatives
\begin{eqnarray}
D_\mu H &=& \partial_\mu H - i \frac{g_Y}{2}W_\mu^a \sigma^a H - i \frac{g_W}{2}B_\mu H,   \nonumber\ \\
D_\mu \Delta &=& \partial_\mu \Delta - i \frac{g_Y}{2}\mathrm{Tr}\left[W_\mu ^a \sigma ^a , \Delta \right] - i \frac{g_W}{2} B_\mu \Delta.
\end{eqnarray}
Here $\sigma^a$ ($a$ = 1, 2, 3) are the Pauli spin matrices and $g_W$ and $g_Y$ are the gauge couplings associated with $SU(2)_L$ and $U(1)_Y$ gauge group respectively. 

%The most general renormalizable scalar tree level potential is given by
%\begin{eqnarray}
%V(H	, \Delta) &=& - m_H^2 \left(H^\dag H\right) + M^2 Tr \left[\Delta^\dag \Delta\right] + \left( \mu_\Delta H^T i \sigma^2 \Delta^+ H +h.c  %\right) + \lambda \left(H^\dag H \right)^2 + \lambda_1 \left(H ^\dag H \right) Tr \left[\Delta^\dag \Delta\right] \nonumber\\\
%&+&  \lambda_2  \left(Tr [\Delta^\dag \Delta]\right)^2 + \lambda_3 Tr \left[(\Delta^\dag \Delta)^2\right] + \lambda_4 \left(H^\dag \Delta %\Delta^\dag H\right).%
%\end{eqnarray}
%Here the trilinear coupling $\mu_\Delta$ can be taken as positive by absorbing its phase into $H $ and $\Delta$.
%=======================================
\paragraph{\underline{Calculation}:}
%==========================================
The dimension six SMEFT operators which contribute Higgs mass correction either 
at one-loop or two-loop level can be written up to a total derivative as, 
\begin{eqnarray}
%\mathcal{Q}_H &=& (H^+ H)^3   \nonumber\ \\
 \mathcal{Q}_{HD} &=& (H^+\mathcal{D}_\mu H)^\ast(H^+\mathcal{D}^\mu H) \supset  \left(\partial_\mu H^\dag\right) HH^\dag \left(\partial^\mu H^\dag\right)+\Big[ \frac{g_W^2}{4}\sigma^a \sigma^b H^\dag  W_\mu^a H H^\dag W^{\mu b}H \nonumber\ \\
&+& \frac{g_Y^2}{4} H^\dag  B_\mu H H^\dag B^{\mu }H \Big ]  \nonumber\ \\
\mathcal{Q}_{H\square} &=& (H^+ H)\square (H^+ H) = - \partial_\mu \left(  H^\dag H \right) \partial^\mu \left( H^\dag H \right)\nonumber\ \\ 
 \mathcal{Q}_{HW} &=& (H^+ H)W_{\mu \nu}^aW^{a,\mu \nu}   \supset  2 H^\dag\Big[ \sigma^a\left( \partial_\mu W_\nu^a \right)\sigma^b \left( \partial^\mu W^{\nu b} \right) - \sigma^a\left( \partial_\mu W_\nu^a \right)\sigma^b \left( \partial^\nu W^{\mu b} \right) \Big] H \nonumber\ \  \\
 &+& g_W^2 \sigma^a f_{abc}\sigma^p f_{pqr}H^\dag W_\mu ^b W_\nu^c W^{\mu q}W^{\nu r} H \nonumber\ \\ 
 \mathcal{Q}_{HB} &=& (H^+ H) B_{\mu\nu}B^{\mu\nu} \supset 2 H^\dag \Big[\partial_\mu B_\nu \partial^\mu B^\nu - \partial_\mu B_\nu \partial^\nu B^\mu   \Big] H \nonumber\ \\
 \mathcal{Q}_{H} &=& (H^\dag H)^3 .
\end{eqnarray}
This result is given in WARSAW basis.
Note that, only momentum dependent vertices can generate quartic divergence at one-loop level. 
Possible Feynman diagrams originating from these terms are similar to Ref.\cite{Biswas:2020abl}.

The authors in Reference \cite{Biswas:2020abl} has performed the calculation in the HISZ basis. 
The transformations between the HISZ and WARSAW bases are given by,
\begin{eqnarray}
\mathcal{Q}_{H D} &=& \mathcal{O}_{\phi,1},\,\,\ \mathcal{Q}_{H \square} = -2 \mathcal{O}_{\phi,2},\,\,\,\ \mathcal{Q}_{H} = 3 \mathcal{O}_{\phi,3},\,\,\ \mathcal{Q}_{GG} = \mathcal{O}_{GG}  \nonumber\ \\
\mathcal{Q}_{H W} &=& \frac{-4}{g_W^2}\mathcal{O}_{WW},\,\,\ \mathcal{Q}_{H B} = \frac{-4}{g_Y^2}\mathcal{O}_{BB},\,\,\  \mathcal{Q}_{HWB} = \frac{-4}{g_W g_Y}\mathcal{O}_{BW}.
\end{eqnarray}
A complete list of transformations can be obtained from Ref:\cite{Brivio:2021alv}. The relevant operators in the 
HISZ basis, $\{\mathcal{O}_{\phi,1}, \mathcal{O}_{\phi,2}, \mathcal{O}_{\phi,3}, \mathcal{O}_{GG}, \mathcal{O}_{WW}, \mathcal{O}_{BB}, \mathcal{O}_{BW} \}$ 
are given in Ref. \cite{Biswas:2020abl,Brivio:2021alv}. It is worth noting that the sign of the coefficients of these SMEFT operators is not the same in both bases. 

We would also like to mention that the result obtained in Eqs. 12 and 13 of Ref: \cite{Biswas:2020abl} can be mapped exactly to our result 
of $\delta m_h^2$, subject to the fact that no operator in the WARSAW basis transforms to the operator 
$\mathcal{O}_{\phi,4}$ ~\cite{Willenbrock:2014bja} in the HISZ basis and there is no contribution of the 
gluonic operator in our model. Due to the above mentioned reasons the parameter space of Wilson coefficients 
found in \cite{Biswas:2020abl} (HISZ basis) and this paper (WARSAW basis) are different.
%===========================================================================================================

\end{document}